\documentclass[12pt]{iopart}
\usepackage{iopams}  
\usepackage{graphicx}
\usepackage{fix-cm}
\usepackage{graphicx}

\newcommand{\ld}{\lambda \Delta}
\newcommand{\ls}{\lambda s}

\begin{document}

\title{Delayed feedback causes non-Markovian behavior 
of neuronal firing statistics}

\author{Kravchuk K.G.  and  Vidybida A.K.}

\address{
Bogolyubov Institute for Theoretical Physics,
Metrologichna str., 14-B, 03680 Kyiv, Ukraine
}
\ead{vidybida@bitp.kiev.ua}
\begin{abstract}

The instantaneous state of a neural network consists of both the
degree of excitation of each neuron, the network is composed of, and positions of impulses in communication lines
between neurons.
In neurophysiological experiments, the neuronal firing moments are registered, but not the state of communication lines.
But  future spiking moments depend essentially on the past positions of
impulses in the lines.
This suggests, that the sequence of intervals between firing moments (interspike intervals, ISIs)
in the network could be non-Markovian.

In this paper, we address this question for a simplest possible neural ``net'', namely,
a single neuron with delayed feedback. The neuron
 receives excitatory input both from the driving Poisson stream and from its own
output through the feedback line.
We obtain analytical expressions for conditional probability density $P(t_{n+1}\mid t_{n},\ldots,t_1,t_{0})$, 
which gives the probability to get an output
ISI of duration $t_{n+1}$ provided the previous $(n+1)$
 output ISIs had durations $t_{n},\ldots,t_1,t_{0}$.
It is proven exactly, that  $P(t_{n+1}\mid t_{n},\ldots,t_1,t_{0})$ does not
reduce to $P(t_{n+1}\mid t_{n},\ldots,t_{1})$ for any $n\ge0$. 
This means that the output ISIs stream cannot be represented as Markov chain of any finite order.

\end{abstract}

\pacs{87.19.ll, 87.10.-e, 02.50.Cw, 02.50.Ey, 87.10.Ca, 87.10.Mn}


\submitto{\JPA}

\section{Introduction}
\label{intro}

In a biological network, the main component parts are neurons and interneuronal 
communication lines -- axons  
\cite{Nicholls}.
 These same units are the main ones in most types of artificial neural networks \cite{Adeli}.
If so, then the instantaneous dynamical state of a network must include dynamical states of all neurons and
communication lines the network is composed of. 
The state of a neuron can be described as its degree of 
excitation.
The state of a line consists of information of
whether the line is empty or conducts an impulse. If it does conduct, then further information about how
much time is required for the impulse to reach the end of the line (time to live) describes the line's state. 

In neurophysiological experiments, the triggering (spiking, firing) moments of individual neurons are registered.
The sequence of intervals between the consecutive moments (interspike intervals, ISIs) is frequently
considered as renewal stochastic process. Recently, based on experimental data 
it was offered that the ISIs sequence could be Markovian of order 4 or higher \cite{RatnamNelson}.

The presence of memory in the ISI sequence is not surprising, taking into account that information about
triggering moments leaves unknown the states of communication lines at those moments. On the other hand,
it is namely the impulses propagating in the communication lines that connect past
firing moments with
the future ones in a reverberating neural network. Without knowledge of
communication line states, information about
previous neuronal firing moments
could improve our predicting ability of the next ones. The exact answer of what kind of memory
could be expected in an ISI sequence of a neuron embedded in a reverberating neural network 
driven with some noisy stimulation requires rigorous mathematical treatment.

\begin{figure}
\includegraphics[width=0.75\textwidth]{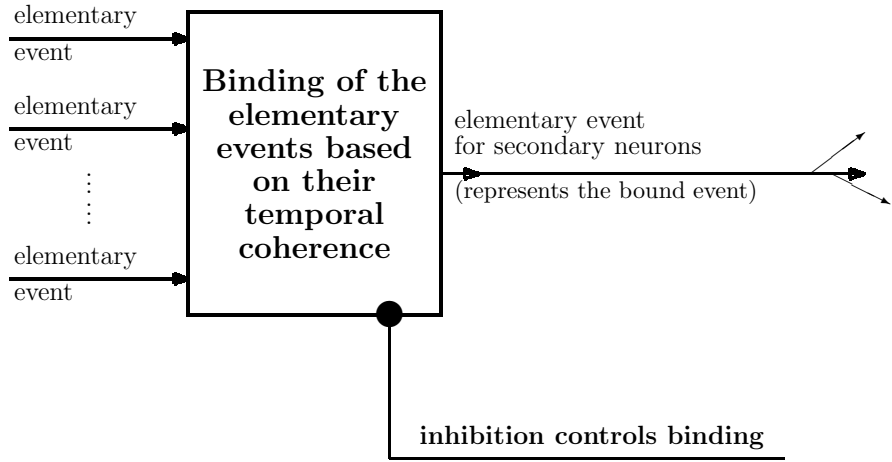}
\caption{Signal processing in the binding neuron model \cite{Vid98}.}
\label{fig:paradigm}
\end{figure}

In this paper, we consider a simplest neural ``net'', namely, a single neuron
with delayed feedback,
which is driven with Poisson process. As neuronal model we take binding neuron
as it allows rigorous mathematical treatment.
We study the ISI output stream of this system and prove that it
cannot be presented as Markovian chain of any finite order. This suggests that activity of a more elaborate
network, if presented in terms of neuronal firing moments, should be non-Markovian as well.


\section{The object under consideration}
\subsection{Binding neuron model}
\label{sec:BN}
The understanding of mechanisms of higher brain functions expects a continuous reduction from higher activities to lower ones, eventually, to activities in individual neurons, 
expressed in terms of membrane potentials and ionic currents. While this approach is correct scientifically and desirable for applications, the complete range of the reduction is unavailable to a single researcher/engineer due to human brain limited capacity. In this connection, it would be helpful to abstract from the rules by which a neuron changes its membrane potentials to rules by which the input impulse signals are processed in the neuron. The “coincidence detector”, and “temporal integrator” are the examples of such an abstraction, 
see discussion in \cite{Konig}.

\begin{figure}
\includegraphics[width=0.75\textwidth]{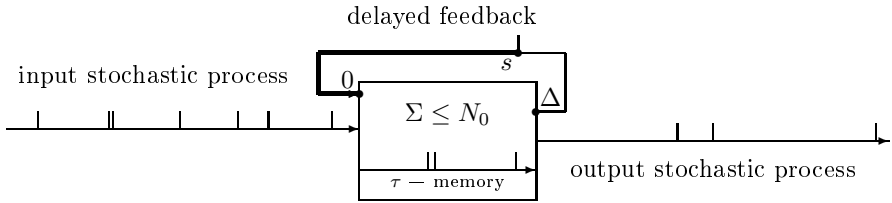}
\caption{Binding neuron with feedback line under Poisson stimulation. 
Multiple input lines with Poisson streams are joined into a single one here.
$\Delta$ is the delay duration in the feedback line.}
\label{fig:BNDF}
\end{figure}

One more abstraction, the binding neuron (BN) model, is proposed as signal processing unit \cite{Vid96}, 
which can operate either as coincidence detector, or temporal integrator, depending on quantitative characteristics of stimulation applied. This conforms with behavior of real neurons, see, e.g. \cite{Rudolph}.
The BN model 
describes functioning of a neuron in terms of discret events, which are input and output impulses, and degree of temporal coherence between the input events, see Figure \ref{fig:paradigm}. Mathematically, this is realized as follows. We expect that all input 
impulses in all input lines are identical.
Each input impulse is stored in the BN for a fixed time, $\tau$. The $\tau$ is similar to the “tolerance interval” discussed in \cite{MacKay}. All input lines are excitatory.  
The neuron fires an output impulse if the number of stored impulses, $\Sigma$, is equal or higher than threshold value, $N_{0}$. 
After that, BN clears its memory and is ready to receive fresh inputs.
That is, every input impulse either disappears contributing to a triggering event, 
or is lost
after spending $\tau$ units of time in the neuron's internal memory.
It is clear, that BN fires when a bunch of input impulses is received in a narrow temporal interval. In this case the bunch could be considered as compound event, and the output impulse --- as an abstract representation of this compound event. One could treat this mechanism as binding of individual input events into a single output event, provided the input events are coherent in time. Such interpretation is suggested by binding of features/events in largescale neuronal circuits 
\cite{Damasio89,Eckhorn,Engel91a}. 

Further, we expect that input stream in each input line is the
Poisson one with some intensity $\lambda_i$. In this case, all input
lines can be collapsed into a single one delivering Poisson stream
of intensity $\lambda=\sum_i\lambda_i$, see Figure \ref{fig:BNDF}.

For analytical derivation, we use BN with $N_0=2$.
The case of higher threshold is considered numerically.

\subsection{Feedback line action}
\label{sec:feedback}
In real neuronal systems, a neuron can have synaptic connection of its axonal branch at its own dendritic tree, see \cite{Aron,Nicoll} for experimental evidence.
As a result, the neuron stimulates itself obtaining excitatory impulse after each firing 
with some propagation delay. We model this situation assuming that
 output impulses of BN  are fed back into BN's input with delay 
$\Delta$. This gives BN with delayed feedback, Figure~\ref{fig:BNDF}.
See also Supplementary Matherial for animation of BN with delayed feedback in action.
Impulses from the feedback line have the same excitatory action on BN as those arrived from Poisson stream. Namely, each one of them is stored in BN's 
memory for time $\tau$, after which it desappears completely, see section \ref{sec:BN}.

The feedback line either keeps one impulse, or keeps no impulses and cannot convey two or more impulses at the same time. If the feedback line is empty at the moment of firing, 
the output impulse enters the line, and after time interval equal $\Delta$ 
reaches the BN's
input. If the line already keeps one impulse at the moment of 
firing,  the just fired impulse ignores the line.

Any output impulse of BN with feedback line may be produced either with impulse from the line involved, or not. We assume that, just after firing and sending output impulse, the line is never empty. This assumption is selfevident for output impulses produced without impulse from the line, or if the impulse from the line was involved, but entered empty neuron. In the letter case, the second (triggering) impulse comes from the Poisson stream, neuron fires and output impulse goes out as well as enters the empty line. On the other hand, if impulse from the line triggers BN, which already keeps one impulse from the input stream, it may be questionable if the output impulse is able to enter the line, 
which was just filled with another impulse. We expect it does. This means that the refraction time of biological axon modelled as feedback line
is equal $\Delta$.
Thus, at the beginning of any output ISI, the line keeps impulse with time to 
live $s$, where $s\in]0;\Delta]$.
In this paper, we consider the case
\begin{equation}\label{case}\Delta<\tau\end{equation}
in order to keep expressions shorter.


\section{Statement of the problem}
\label{sec:problem}

The input stream of impulses, which drives neuronal activity is stochastic.
Therefore, the output activity of our system requires probabilistic description
in spite of the fact that both the BN and the feedback line action mechanisms
are deterministic.
We treat the output stream of BN with delayed feedback as the 
stationary process\footnote{
The stationarity of the output stream results both from the stationarity of the input 
one and from 
the absence of adaptation in the BN model, see Section~\ref{sec:BN}. 
In order to ensure stationarity, we also expect that system is considered after initial period sufficient to forget the initial conditions.
}.
In order to discribe its statistics, we introduce the following basic functions:
\begin{itemize}
	\item joint probability density $P(t_{m},t_{m-1},\ldots,t_{0})$ for $(m+1)$
successive output ISI durations.
        \item conditional probability density $P(t_{m}\mid t_{m-1},\ldots,t_{0})$ for output ISI durations; $P(t_{m}\mid t_{m-1},\ldots,t_{0})\rmd t_{m}$ gives the probability to obtain an output ISI of duration between $t_{m}$ and $t_{m}+\rmd t_{m}$ provided previous $m$ ISIs had durations $t_{m-1},t_{m-2},\ldots,t_{0}$, respectively.
\end{itemize}

\begin{description}
\item[Definition]
\it
The sequence of random variables $\{t_{j}\}$, taking values in $\Omega$, is called the Markov chain of the order $n\ge0$, if 
\begin{equation*}
	\forall_{m > n} \forall_{t_0\in\Omega}\ldots \forall_{t_m\in\Omega}\  
        P(t_{m}\mid t_{m-1},\ldots,t_{0}) 
	= P(t_{m}\mid t_{m-1},\ldots,t_{m-n}),
\end{equation*}
and this equation does not hold for any $n'<n$ (e.g. \cite{Doob}).
In the case of ISIs one reads $\Omega=\mathbb{R^+}$.
\end{description}

In particular, taking $m=n+1$, we have the necessary condition 
\begin{equation}\label{def}\fl
	P(t_{n+1}\mid t_{n},\ldots,t_{1},t_{0}) 
	= P(t_{n+1}\mid t_{n},\ldots,t_{1}),
	\qquad t_i\in\Omega, \qquad i=0,\ldots,n+1,
\end{equation}
required for the stochastic process $\{t_{j}\}$ to be the $n$-order Markov chain.

\begin{description}
\item[Theorem 1 \label{theo}]
\emph{The output ISIs stream of BN with delayed feedback under Poisson stimulation
cannot be represented as a Markov chain of any finite order.}
\end{description}

\section{Proof outline}
\label{sec:outline}
In order to prove the Theorem 1, 
we are going to show analytically, that 
the equality (\ref{def}) does not hold for any finite value of $n$, namely,
in the exact expression for conditional probability density $P(t_{n+1}\mid t_{n},\ldots,t_{1},t_{0})$, 
elimination of $t_{0}$-dependence is impossible. 

For this purpose we introduce the stream of events $(t,s)$
\begin{displaymath}
	\mathbf{ts} = \{\dots, (t_i,s_i),\dots\},
\end{displaymath}
where $s_i$ is the time to live of the impulse in the feedback line at the moment, when ISI $t_i$ starts.
We consider the joint probability density $P(t_{n+1},s_{n+1};t_{n},s_{n};\ldots;t_{0},s_{0})$ for realization of $(n+2)$ successive events $(t,s)$, and the corresponding conditional probability density $P(t_{n+1},s_{n+1}\mid t_{n},s_{n};\ldots;t_{0},s_{0})$ for these events.

\begin{description}
\label{lemma}
\item[Lemma 1]
\emph{
Stream $\mathbf{ts}$ is 1-st order markovian:
\begin{eqnarray}
\fl
\nonumber
	\forall_{n\ge0}
	\forall_{t_{0}>0} \forall_{s_0\in]0;\Delta]}\ldots
	\forall_{t_{n+1}>0} \forall_{s_{n+1}\in]0;\Delta]}
\\ 
\label{marka}
	P(t_{n+1},s_{n+1}\mid t_{n},s_{n};\ldots;t_{0},s_{0}) 
	= P(t_{n+1},s_{n+1}\mid t_{n},s_{n}),
\end{eqnarray}
where $\{t_0,\ldots,t_{n+1}\}$ is the set of successive ISIs,
and $\{s_0,\ldots,s_{n+1}\}$ are corresponding times to live.
}
\end{description}

\begin{description}
\item[Proof]
	Indeed, the value of $s_{n+1}$ characterizes the state of the system at the moment 
        of triggering, $\theta$, and the value of $t_{n+1}$ characterizes the system's behavior after
        that triggering, which means that, in physical time, $s_{n+1}$ always gets its value before than the $t_{n+1}$ does. Once the value of $s_{n+1}$ is known, the realization of $t_{n+1}$ is completely determined by a unique
realization  of the input Poisson process after the $\theta$.

	At the same time, in $P(t_{n+1},s_{n+1}\mid t_{n},s_{n},\ldots,t_{0},s_{0})$ the value of $s_{n+1}$ can be derived unambiguously from $(t_{n},s_{n})$ (See 		Sections~\ref{sec:feedback} and \ref{sec:Pts|ts}):
\begin{eqnarray}
\nonumber
		s_{n+1} &=  s_n - t_n, \qquad &t_n < s_n, 
\\
		&=\Delta, \qquad &t_n \ge s_n.
\end{eqnarray}
	Just after triggering, BN appears in the standard state 
(it is empty), the state of line is given by the value of 
$s_{n+1}$, and the state of input Poisson stream is always the same.
Therefore, once the pair of values $(t_n,s_n)$ is given, the state of the system at the moment of $(n+1)$-th ISI beginning is determined completely, and knowledge of previous values of $(t_i,s_i),\, i<n$ 
adds nothing to our predictive ability as regards the values of 
$(t_{n+1},s_{n+1})$, which proves (\ref{marka}).
\end{description}

In order to find the conditional probability density $P(t_{n+1}\mid t_{n},\ldots,t_{1},t_{0})$, the following steps should be performed:
\begin{itemize}
	\item {\em Step 1.} Use property \eref{marka} for calculating joint probability of events $(t,s)$:
\begin{eqnarray}
\fl
	P(t_{n+1},s_{n+1};t_{n},s_{n};\ldots;t_{0},s_{0}) = \nonumber
\\
\label{mark}
	P(t_{n+1},s_{n+1}\mid t_{n},s_{n}) 
	\ldots P(t_{1},s_{1}\mid t_{0},s_{0}) P(t_{0},s_{0}),
\end{eqnarray}
where $P(t,s)$ and $P(t_{n},s_{n}\mid t_{n-1},s_{n-1})$ denote the stationary
probability density and conditional probability density (transition probability) 
for events $(t,s)$.

	\item {\em Step 2.} Represent the joint probability density for successive output ISI durations as marginal probability
by integration over variables $s_i,\, i=0,1,\dots,n+1$:
\begin{eqnarray}
\fl
	P(t_{n+1},t_{n},\ldots,t_{0}) = \nonumber
\\
\label{main}
	\int_{0}^{\Delta}\rmd s_{0} 
	\int_{0}^{\Delta}\rmd s_{1} \ldots \int_{0}^{\Delta}\rmd s_{n+1} 
	P(t_{n+1},s_{n+1};t_{n},s_{n};\ldots;t_{0},s_{0}).
\end{eqnarray}

	\item {\em Step 3.} Use the definition of conditional probability density:
\begin{equation}
\label{defcond}
	P(t_{n+1}\mid t_{n},\ldots,t_{1},t_{0})
	= \frac{P(t_{n+1},t_{n},\ldots,t_{0})}
	{P(t_{n},\ldots,t_{0})}.
\end{equation}

\end{itemize}

Taking into account Steps 1 and 2, for joint probability density $P(t_{n+1},\ldots,t_{0})$ one derives 
\begin{eqnarray}
\fl
	P(t_{n+1},t_{n},\ldots,t_{0}) = \nonumber
\\
\label{joint}
	\int_{0}^{\Delta}\rmd s_{0} \ldots \int_{0}^{\Delta} \rmd s_{n+1}
	P(t_{0},s_{0})\ 
	\prod_{k=1}^{n+1} P(t_{k},s_{k}\mid t_{k-1},s_{k-1}).
\end{eqnarray}

In the next section, we are going to find the exact analytical expressions for probability densities $P(t,s)$ and $P(t_k,s_k\mid t_{k-1},s_{k-1})$, 
and perform the integration in (\ref{joint}). Then we aply the Step 3, 
above, to find expressions for conditional probabilities 
$P(t_{n+1}\mid t_{n},\ldots,t_{1},t_{0})$. It appears, that the conditional probabilities have singular parts of the Dirac's $\delta$-function type. This is because the system's dynamics involves discret
events of obtaining impulse by neuron (see below).
In order to prove that the equality~(\ref{def}) does not hold for any
$n\ge0$, we use the singular parts only.

\section{Main calculations}
\label{sec:main}
\subsection{Probability density $P(t,s)$ for events $(t,s)$}
\label{sec:Pts}
The probability density $P(t,s)$ can be derived as the product
\begin{equation}
\label{P(t,s)}
	P(t,s) = F(t\mid s) f(s),	
\end{equation}
where $f(s)$ denotes the stationary probability density for time to live of the impulse in the feedback line at the moment of an output ISI beginning, 
$F(t\mid s)$ denotes conditional
probability density for ISI duration provided the time to live of the impulse in the feedback line equals $s$ at the moment of this ISI beginning. 
Exact expressions for both $f(s)$ and $F(t\mid s)$ are given in  \cite[Eqs.(5),(6) and (31)]{BNDF}.
In this paper we need only singular parts of those expressions, which read:
\begin{equation}
\label{P(t|s)sing}
	F^{\textrm{sing}}(t\mid s) = \ls \rme^{-\ls} \delta(t-s),	
\end{equation}
\begin{equation}
\label{fsing}
	f^{\textrm{sing}}(s) = a \cdot \delta(s-\Delta),\qquad\textrm{where}\qquad a=\frac{4\rme^{2\ld}}{(3+2\ld)\rme^{2\ld}+1},
\end{equation}
where  $a$ gives the probability to obtain the impulse in the feedback line with time to live equal $\Delta$ at the beginning of an arbitrary ISI, $\lambda$ --- is the input Poisson stream intensity.

\begin{figure}
	\includegraphics[width=0.5\textwidth,angle=0]{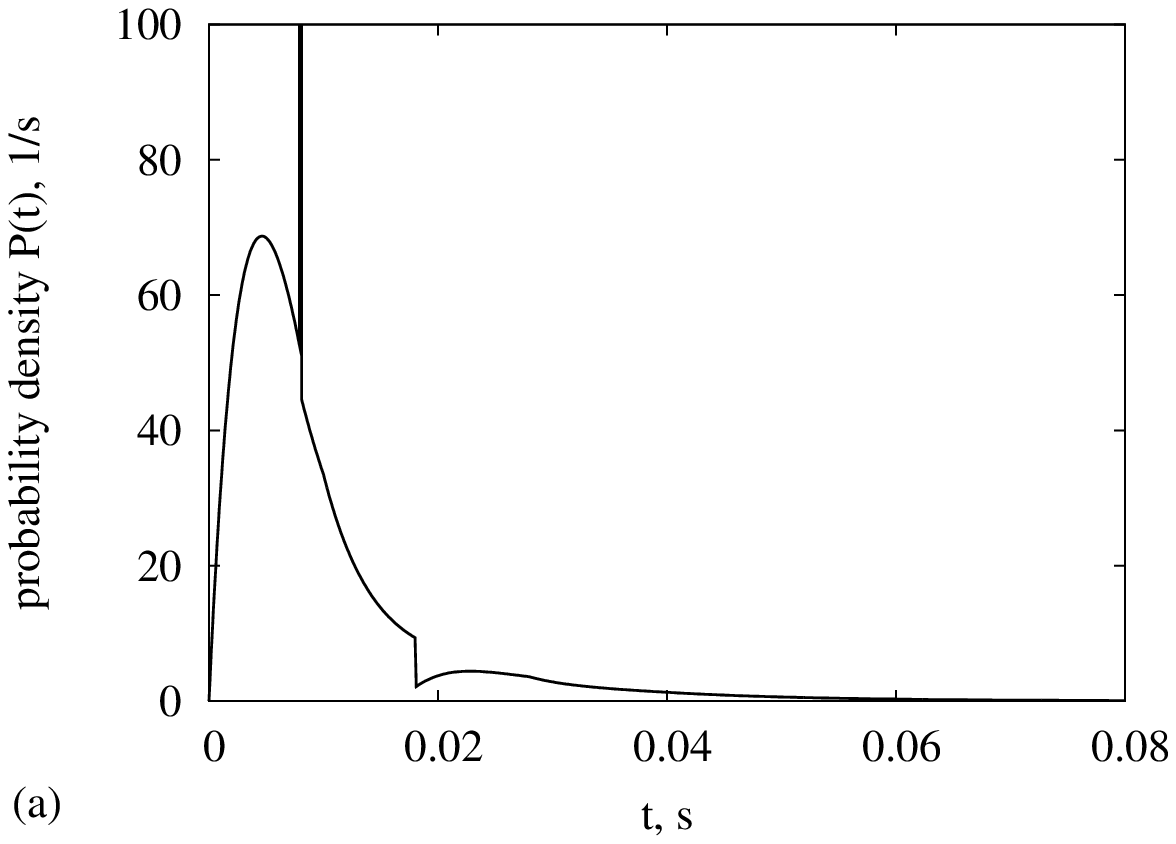}
	\includegraphics[width=0.5\textwidth,angle=0]{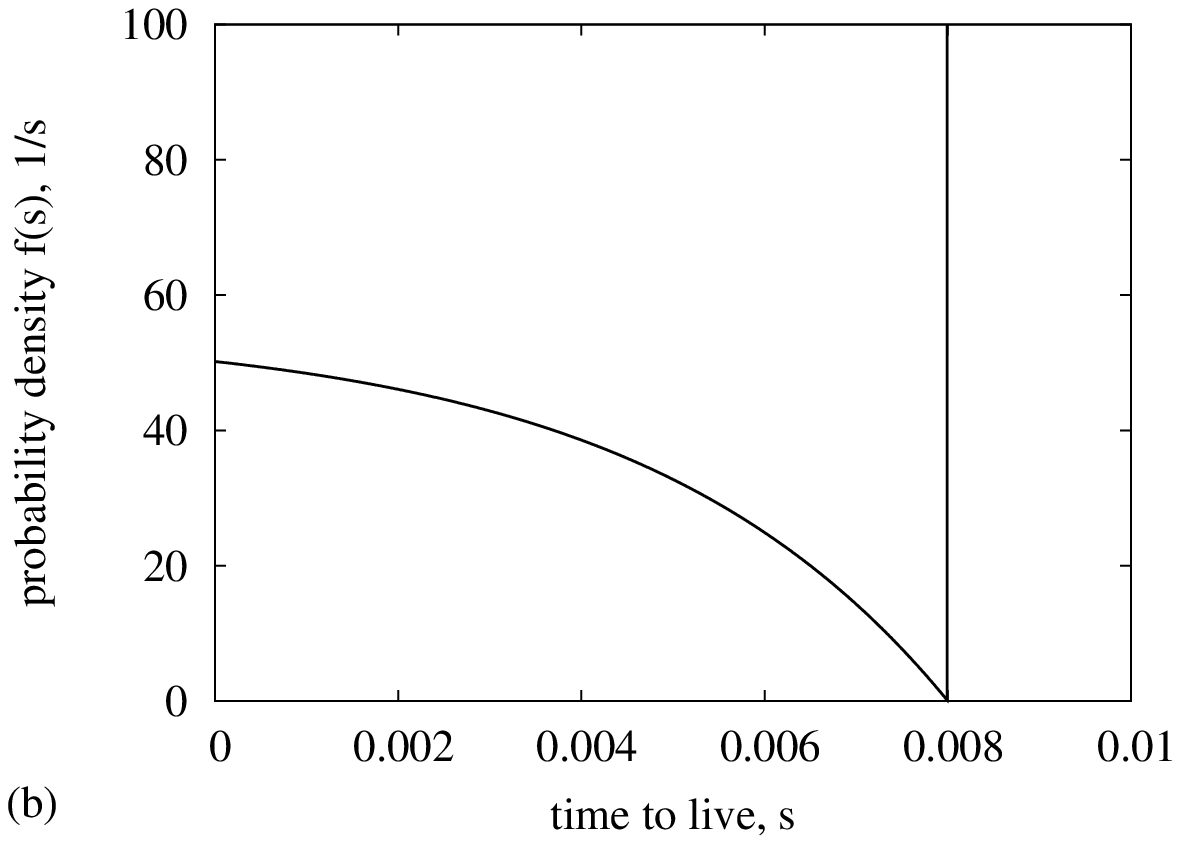}
\caption{Output ISI probability density $P(t)$ (a) and probability density $f(s)$  
for times to live of the impulse in the feedback line (b), found analytically in \cite{BNDF}.
Here $\tau$ = 10 ms, 
$\Delta$ = 8 ms, $\lambda$ = 150 s$^{-1}$, $N_0$=2. The presence of $\delta$-function
in both densities is clearly visible.}
\label{fig:Ptfs}       
\end{figure}

The presence of $\delta$-functions in $F(t\mid s)$ and $f(s)$ can be explained as follows. The probability to obtain an output ISI of duration $t$ exactly equal $s$ is not infinitesimally 
small. Due to (\ref{case}),  it equals to
the probability to obtain exactly one impulse from the Poisson stream during time interval $]0;s[$, which is $\ls \rme^{-\ls}$. 
The second impulse comes from the line
and triggers the neuron exactly after time interval $s$.
So, we have the non-zero probability to obtain an output ISI of duration exactly equal to $s$.
This gives the $\delta$-function at $t=s$ in the probability density 
$F(t\mid s)$.

The probability to have time to live, $s$, exactly equal
$\Delta$ at the moment of an output ISI beginning is not infinitessimally 
small as well. 
Every time, when the line is free at the moment of an output ISI beginning, the impulse enters the line and has time to live equal $\Delta$. For the line to be free from impulses at the moment of triggering, it is nessesary that $t\ge s$ for the previous ISI. The set of realizations of the input Poisson process, each realization
results in $t\ge s$, has non-zero probability $a$, see \eref{fsing},
and this gives the $\delta$-function at $s=\Delta$ in the probability density 
$f(s)$.

The output ISI probability density $P(t)$ can be obtained as the result of integration of (\ref{P(t,s)}) (see \cite{BNDF} for details):
\begin{equation}
\label{P(t)}
	P(t) = \int_{0}^{\Delta} F(t|s)f(s)\rmd s.
\end{equation}

Examples of $P(t)$ and $f(s)$ graphs are given in Figure~\ref{fig:Ptfs}.

\subsection{Conditional probability density $P(t_k,s_k\mid t_{k-1},s_{k-1})$}
\label{sec:Pts|ts}
Here we find the conditional probability density 
$P(t_k,s_k\mid t_{k-1},s_{k-1})$ 
for events $(t_k,s_k)$, which determines the probability to obtain the event
$(t_k,s_k)$, with precision $\rmd t_k \rmd s_k$,
provided the previous event was $(t_{k-1},s_{k-1})$.
By definition of conditional probabilities, the probability density wanted can be represented as the following product
\begin{equation}
	P(t_k,s_k\mid t_{k-1},s_{k-1}) = 
	F(t_k\mid s_k,t_{k-1},s_{k-1}) f(s_k\mid t_{k-1},s_{k-1}),
\end{equation}
where $F(t_k\mid s_k,t_{k-1},s_{k-1})$ denotes conditional probability
density for ISI duration, $t_k$, 
provided i) this ISI started with lifetime of impulse in the feedback line
equal to $s_k$, and ii) previous $(t,s)$-event was $(t_{k-1},s_{k-1})$; 
$f(s_k\mid t_{k-1},s_{k-1})$ denotes conditional probability density 
for times to live of the impulse in the feedback line under condition ii). 
It is obvious, that
\begin{equation}
	F(t_k\mid s_k,t_{k-1},s_{k-1}) = F(t_k\mid s_k),
\end{equation}
because with $s_k$ being known, the previous event $(t_{k-1},s_{k-1})$ does not add any information, useful to predict $t_k$ (compare with proof
of Lemma 1).

In order to find the probability density $f(s_k\mid t_{k-1},s_{k-1})$, let us consider different relations between $t_{k-1}$ and $s_{k-1}$. If $t_{k-1} \ge s_{k-1}$, the line will have time to get free from the impulse during the ISI $t_{k-1}$. That is why at the beginning of ISI $t_k$, an output spike will enter the line and will have time to live equal $s_k=\Delta$ with 
 probability 1. 
 Therefore, the probability density contains the corresponding delta-function:
\begin{equation}
	f(s_k\mid t_{k-1},s_{k-1}) = \delta(s_k-\Delta), \qquad t_{k-1}\ge s_{k-1}.
\end{equation}
If $t_{k-1} < s_{k-1}$, than the ISI $t_{k-1}$ ends before the impulse leaves the feedback line. Therefore, at the beginning of the $t_k$,
the line still keeps the same impulse as at the beginning of $t_{k-1}$. This impulse has time to live being accurately equal to 
$s_k = s_{k-1}-t_{k-1}$, so
\begin{equation}
	f(s_k\mid t_{k-1},s_{k-1}) =
	\delta(s_k-s_{k-1}+t_{k-1}), \qquad t_{k-1} < s_{k-1}.	
\end{equation}
Taking all together, for the conditional probability density $P(t_k,s_k\mid t_{k-1},s_{k-1})$ one obtains
\begin{eqnarray}
\nonumber
	P(t_k,s_k\mid t_{k-1},s_{k-1})
	&=F(t_k\mid s_k)\delta(s_k-\Delta), \qquad\qquad&t_{k-1}\ge s_{k-1},
\\
\label{P(t,s|t,s)}
	&=F(t_k\mid s_k)\delta(s_k-s_{k-1}+t_{k-1}), \qquad &t_{k-1} < s_{k-1},
\end{eqnarray}
where exact expression for $F(t\mid s)$ is given in  \cite[Eqs.(5),(6)]{BNDF}.



\subsection{Joint probability density $P(t_{n+1},\ldots,t_{0})$}
\label{sec:joint}
In this section, we are going to find the exact analytical expression for the joint probability density $P(t_{n+1},\ldots,t_{0})$ 
at the domain 
\begin{equation}
\label{domain}
	D_1=\left\{(t_0,\ldots,t_n)\ \ \Big |\ \sum_{i=0}^{n}t_{i}<\Delta\right\}.
\end{equation}
It is worth to notice, that the set of $(n+1)$ 
successive ISI durations $t_0,\ldots,t_n$ has non-zero probability, $p_\Delta>0$, to fall into the domain (\ref{domain}). Indeed, BN with threshold $N_{0}=2$ needs $2(n+1)$ input impulses within time window $]0;\Delta[$ to be triggered $(n+1)$
 times within this window (condition 
(\ref{case}) ensures that no input impulse is lost). BN receives impulses both from the Poisson stream and from the line. 
But no more than one impulse from the line may have time to reach BN's input during time interval less than $\Delta$. Therefore, the other $(2n+1)$
 impulses must be received from the Poisson stream. On the other hand, if $2(n+1)$ input impulses are received from the Poisson stream during time interval $]0;\Delta[$, the inequality 
(\ref{domain}) holds for sure, no matter is the impulse from the feedback line involved, or not. 
Therefore,
$
	p_{\Delta} > p(2n+2,\Delta) > 0,
$
where $p(i,\Delta)$ gives the probability to obtain $i$ impulses from the Poisson stream 
during time interval $\Delta$ \cite{Feller}: 
$p(i,\Delta) = \rme^{-\ld}{(\ld)^{i}}/{i!}$. 

Having in mind (\ref{domain}),
let us split the integration domain for $s_{0}$ in (\ref{joint}) in the following way: 
\begin{equation*}
	 \int_{0}^{\Delta}\rmd s_{0}
	= \int_{0}^{t_{0}}\rmd s_{0} 
	+ \sum_{i=1}^{n}\int_{\sum_{j=0}^{i-1}t_{j}}^{\sum_{j=0}^{i}t_{j}}\rmd s_{0}
	+ \int_{\sum_{j=0}^{n}t_{j}}^{\Delta}\rmd s_{0},
\end{equation*}
and introduce the notations:
\begin{eqnarray}
\fl\nonumber
	 I_{i} = \int\limits_{\sum_{j=0}^{i-1}t_{j}}^{\sum_{j=0}^{i}t_{j}}\rmd s_{0}
	\int\limits_{0}^{\Delta}\rmd s_{1} \ldots \int\limits_{0}^{\Delta} \rmd s_{n+1}
	P(t_{0},s_{0})\ 
	\prod_{k=1}^{n+1} P(t_{k},s_{k}\mid &t_{k-1},s_{k-1}),
\\
\label{Iidef}
	&i=0,1,2,\ldots,n,
\end{eqnarray}
\begin{equation}
\label{In+1def}
	 I_{n+1} = 
         \int\limits_{\sum\limits_{j=0}^{n}t_{j}}^{\Delta}\rmd s_{0}
	\int\limits_{0}^{\Delta}\rmd s_{1} \ldots \int\limits_{0}^{\Delta} \rmd s_{n+1}
	P(t_{0},s_{0})\ \prod_{k=1}^{n+1} P(t_{k},s_{k}\mid t_{k-1},s_{k-1}),
\end{equation}
where we assume, that $\sum_{j=j_{1}}^{j_{2}}=0$ for $j_{1}>j_{2}$.

Consider the fixed sequence of ISIs, $(t_0,\ldots,t_n)$,
which belongs to $D_1$. 
Domain of $s_{0}$ values covered by $I_{i}$, $i=0,\ldots,n$,
corresponds to the scenario, when impulse, which was in the feedback line at the 
beginning of interval $t_{0}$ (with time to live $s_0$), will reach BN during 
interval $t_{i}$.
In this process, after each
firing, which starts 
ISI $t_k$, $k\le i$, the time to live of the impulse in the feedback line is decreased
exactly by $t_{k-1}$. This means, that variables of integration $\{s_0,\dots,s_{n+1}\}$, above,
are not actually independent, but must satisfy the following relations:
\begin{equation}
\label{relations1}
	s_k=s_0-\sum\limits_{j=0}^{k-1}t_j,
	\qquad k=1,\dots,i,
\end{equation}
which are ensured by $\delta$-function in the bottom line of (\ref{P(t,s|t,s)}).
Next to $s_i$ time to live must be equal $\Delta$:
\begin{equation}
\label{relation}
	s_{i+1}=\Delta,
\end{equation}
and this is ensured by $\delta$-function in the top line of (\ref{P(t,s|t,s)}).
The next to $s_{i+1}$ times to live again are decreased by corresponding ISI with
each triggering. Due to (\ref{domain}), this brings about 
another set of relations:
\begin{equation}
\label{relations2}
	s_k=\Delta-\sum\limits_{j=i+1}^{k-1}t_j,
	\qquad k=i+2,\dots,n+1,
\end{equation}
which are again ensured by $\delta$-function in the bottom line of (\ref{P(t,s|t,s)}).
Relations (\ref{relations1}), (\ref{relation}) and (\ref{relations2}) together with limits
of integration over $s_0$ in (\ref{Iidef}) ensure that at $D_1$ the following inequalities hold:
\begin{equation}\label{ineqs}
\eqalign{
	s_k>t_k,\qquad k=0,\dots,i-1,
\cr	
	s_i\le t_i,
\cr
	s_k>t_k, \qquad k=i+1,\dots,n.
}
\end{equation}
Inequalities (\ref{ineqs})
allow one to decide
correctly which part of rhs of (\ref{P(t,s|t,s)}) should replace each transition probability
$P(t_{k},s_{k}\mid t_{k-1},s_{k-1})$ in (\ref{Iidef}), and perform all but one 
integration. This gives:

\begin{eqnarray}
\label{Ii}
\fl\nonumber
	 I_i &= \int\limits_{\sum_{j=0}^{i-1}t_j}^{\sum_{j=0}^i t_j} \rmd s_0
	&\int\limits_0^{\Delta}\rmd s_1\cdot\ldots\cdot \int\limits_0^{\Delta} \rmd s_{n+1}
         F(t_0\mid s_0) f(s_0)
        \prod_{k=1}^i  F(t_k\mid s_k) \delta(s_k - s_0 + \sum_{j=0}^{k-1}t_j)
\\\nonumber\fl
 	&\times  F(t_{i+1}\mid s_{i+1}&)\ \delta(s_{i+1} - \Delta)
	\prod_{k=i+2}^{n+1} 
        F(t_k\mid s_k) \delta(s_k-\Delta+\sum_{j=i+1}^{k-1}t_j)
\\\nonumber\fl
	&= F(t_{n+1}\mid \Delta &- \sum_{j=i+1}^n t_j)
	F(t_n\mid \Delta - \sum_{j=i+1}^{n-1} t_j)\cdot\ldots \cdot
	F(t_{i+2}\mid \Delta - t_{i+1})  F(t_{i+1}\mid \Delta) 
\\\nonumber\fl
	&\times\int\limits_{\sum_{j=0}^{i-1}t_j}^{\sum_{j=0}^i t_j} 
	F(t_i \mid &s_0 - \sum_{j=0}^{i-1}t_j)
	F(t_{i-1}\mid s_0 - \sum_{j=0}^{i-2} t_j)\cdot\ldots\cdot
	F(t_1\mid s_0-t_0) F(t_0\mid s_0) f(s_0) \rmd s_0,
\\\fl
	&\qquad&i=0,1,2,\ldots,n.
\end{eqnarray}

The last expression might be obtained as well by means of consecutive substitution
of either top, or bottom line of (\ref{P(t,s|t,s)}) into (\ref{Iidef}), without 
previously discovering (\ref{relations1}) -- (\ref{ineqs}).

Finally, integral $I_{n+1}$ corresponds to the case, when at the beginning of 
interval $t_{n+1}$, the line still keeps the same impulse as at the beginning of $t_{0}$. Therefore, $I_{n+1}$ comprises the rest of scenarios contributing to the value of $P(t_{n+1},\ldots,t_{0})$ in (\ref{main}). 
Here, the bottom line of (\ref{P(t,s|t,s)}) ensures that values of
variables of integration $\{s_0,\dots,s_{n+1}\}$, which contribute to the $I_{n+1}$,
should satisfy the following relations:
\begin{equation}
	s_k=s_0-\sum\limits_{j=0}^{k-1}t_j,
	\qquad k=1,\dots,n+1,
\end{equation}
which taken at the domain $D_1$, defined in (\ref{domain}), results in inequalities
\begin{equation}
\label{ineqsn+1}
	s_k>t_k,
	\qquad k=0,\dots,n.
\end{equation}
Equations (\ref{P(t,s|t,s)}) and (\ref{ineqsn+1}) allow one to perform integration in (\ref{In+1def})
and to obtain:
\begin{eqnarray}\label{In+1}
\fl\nonumber
	 I_{n+1} &= \int_{\sum_{j=0}^{n}t_{j}}^{\Delta}\rmd s_{0}
	\int_{0}^{\Delta}\rmd s_{1} \ldots \int_{0}^{\Delta} \rmd s_{n+1}
 	F(t_{0}\mid s_{0}) f(s_{0})
	\prod_{k=1}^{n+1} F&(t_{k}\mid s_{k}) \delta(s_{k}-s_{0}+\sum_{j=0}^{k-1}t_{j}) 
\\\nonumber\fl
	&= \int\limits_{\sum_{j=0}^{n}t_{j}}^{\Delta} 
	F(t_{n+1}\mid s_{0}-\sum_{j=0}^{n}t_{j})
	F(t_{n}\mid s_{0}-\sum_{j=0}^{n-1}t_{j})\ldots
	F(t_{1}&\mid s_{0}-t_{0})
\\\fl 
        &\qquad&\times F(t_{0}\mid s_{0}) f(s_{0}) \rmd s_{0}.
\end{eqnarray}

Taking into account (\ref{Ii}) and (\ref{In+1}), one obtains the following expression for joint probability density $P(t_{n+1},\ldots,t_{0})$:
\begin{eqnarray}
\fl\nonumber
	P(t_{n+1},\ldots,t_{0}) &=
	\sum_{i=0}^{n+1} I_{i}
\\\nonumber
	&=\sum_{i=0}^{n} F(t_{i+1}\mid \Delta)\  
	\prod_{k=i+2}^{n+1}	
	F(t_{k}\mid \Delta-\sum_{j=i+1}^{k-1} t_{j}) \nonumber
\\
	&\times
	\int_{\sum_{j=0}^{i-1}t_{j}}^{\sum_{j=0}^{i}t_{j}} F(t_{0}\mid s_{0})\ f(s_{0}) 
	\prod_{k=1}^{i}	
	F(t_{k}\mid s_0-\sum_{j=0}^{k-1} t_{j}) \rmd s_{0} \nonumber
\\
\label{P}
	&+
	\int_{\sum_{j=0}^{n}t_{j}}^{\Delta} 
	F(t_{0}\mid s_{0})\ f(s_{0})\ \prod_{k=1}^{n+1}F(t_{k}\mid s_{0}-\sum_{j=0}^{k-1} t_{j})
	\rmd s_{0},
	\qquad \sum_{i=0}^{n}t_{i}<\Delta,
\end{eqnarray}
where we assume, that $\sum_{j=j_{1}}^{j_{2}}=0$ and $\prod_{j=j_{1}}^{j_{2}}=1$ for $j_{1}>j_{2}$.

Using \eref{defcond}, for conditional probability 
density $P(t_{n+1}\mid t_n,\ldots,t_{0})$ one derives:
\begin{eqnarray}
\fl\nonumber
	P(t_{n+1}\mid t_n,\ldots,t_{0}) 
	&=\frac{1}{P(t_n,\ldots,t_{0})}\cdot\Big(
	\sum_{i=0}^{n} F(t_{i+1}\mid \Delta)\  
	\prod_{k=i+2}^{n+1}	
	F(t_{k}\mid \Delta&-\sum_{j=i+1}^{k-1} t_{j})\times \nonumber
\\
	&\times
	\int_{\sum_{j=0}^{i-1}t_{j}}^{\sum_{j=0}^{i}t_{j}} F(t_{0}\mid s_{0})\ f(s_{0}) 
	\prod_{k=1}^{i}	
	F(t_{k}\mid s_0-\sum_{j=0}^{k-1} t_{j}) \rmd &s_{0} \nonumber
\\\nonumber
	&+ 
	\int_{\sum_{j=0}^{n}t_{j}}^{\Delta} 
	F(t_{0}\mid s_{0})\ f(s_{0})\ \prod_{k=1}^{n+1}F(t_{k}\mid s_{0}-\sum_{j=0}^{k-1} t_{j})
	\rmd &s_{0}\Big),
\\
\label{Pcond}
	&\qquad &\sum_{i=0}^{n}t_{i}<\Delta,
\end{eqnarray}
where expression for $P(t_n,\ldots,t_{0})$ can be obtained from \eref{P} with $(n-1)$ 
substituted instead of $n$.



\subsection{Singular part of $P(t_{n+1},\ldots,t_{0})$}
\label{sec:joint_sing}
In order to obtain the singular part
of expression, 
defined in (\ref{P}), let us first derive singular parts for all $I_{i}$, $i=0,\ldots,n$ and $I_{n+1}$ separately. 
In order to keep the expressions shorter, we represent $I_{i}$  as follows
\begin{equation}\label{XY}
	I_{i} (t_{0},\ldots,t_{n+1}) = 
	X_{i}(t_{0},\ldots,t_{i})\cdot Y_{i}(t_{i+1},\ldots,t_{n+1}),
	\qquad i = 0,1,\ldots,n,
\end{equation}
where
\begin{equation}
\label{X}\fl
	X_{i}\equiv
	\int\limits_{\sum_{j=0}^{i-1}t_{j}}^{\sum_{j=0}^{i}t_{j}} 
	F(t_{i}| s_{0}-\sum_{j=0}^{i-1}t_{j})
	F(t_{i-1}| s_{0}-\sum_{j=0}^{i-2}t_{j})\ldots
	F(t_{1}| s_{0}-t_{0}) F(t_{0}| s_{0}) f(s_{0}) \rmd s_{0},
\end{equation}
\begin{equation}
\fl
	Y_{i}\equiv
	F(t_{n+1}\mid \Delta - \sum_{j=i+1}^{n}t_{j})
	F(t_{n}\mid \Delta - \sum_{j=i+1}^{n-1}t_{j})\ldots 
	F(t_{i+2}\mid \Delta-t_{i+1})  F(t_{i+1}\mid \Delta).
\end{equation}
It is clear, that at $D_1$, $X_{i}$ is the part of the probability density for $(i+1)$ successive ISI durations, which corresponds to the case when the impulse, which was in the line at the beginning of the first ISI, reaches the neuron's input within the last one.
And the $Y_{i}$ gives the probability density for $(n+1-i)$ successive ISI durations provided the impulse enters the line just at the beginning of the first one of these ISIs.

At the domain considered, namely, for $\sum_{i=0}^{n}t_{i}<\Delta$, 
the expressions for $F(t_{n}\mid \Delta - \sum_{j=i+1}^{n-1}t_{j})$, \dots, 
$F(t_{i+2}\mid \Delta-t_{i+1})$ and $ F(t_{i+1}\mid \Delta)$ have no singularities,
see (\ref{P(t|s)sing}). Therefore
\begin{equation}
\fl
	Y_{i}^{\textrm{sing}} = F^{\textrm{sing}}(t_{n+1}| \Delta - \sum_{j=i+1}^{n}t_{j})
	F(t_{n}| \Delta - \sum_{j=i+1}^{n-1}t_{j})\ldots 
	F(t_{i+2}| \Delta-t_{i+1})  F(t_{i+1}| \Delta). 
\end{equation}
At the same time, intergation limits in (\ref{X}) ensure that 
$
	X_{i}^{\textrm{sing}}=0.
$
Indeed, each integral $X_{i}$ (and, originally, $I_{i}$), $i=0,1,\ldots,n$, covers the half-open interval $s_{0}\in\big]\sum_{j=0}^{i-1}t_{j};\sum_{j=0}^{i}t_{j}\big]$. The only singularity of integrand in (\ref{X}) at this domain is $\delta(\sum_{j=0}^{i}t_{j}-s_{0})$ provided by $F(t_{i}\mid s_{0}-\sum_{j=0}^{i-1}t_{j})$, see (\ref{P(t|s)sing}), 
and it disappears after intergation.
Therefore
\begin{eqnarray}
\label{Iising}\fl\nonumber
	 I_{i}^{\textrm{sing}} = 
	F^{\textrm{sing}}(t_{n+1}| \Delta - \sum_{j=i+1}^{n}t_{j})
	F(t_{n}| \Delta - \sum_{j=i+1}^{n-1}t_{j})\ldots 
	F(t_{i+2}| \Delta-t_{i+1})  F(t_{i+1}| \Delta)
\\\nonumber\fl
	\times\int_{\sum_{j=0}^{i-1}t_{j}}^{\sum_{j=0}^{i}t_{j}} 
	F(t_{i}\mid s_{0}-\sum_{j=0}^{i-1}t_{j})
	F(t_{i-1}\mid s_{0}-\sum_{j=0}^{i-2}t_{j})\ldots
	F(t_{1}\mid s_{0}-t_{0}) F(t_{0}\mid s_{0}) f(s_{0}) \rmd s_{0},
\\
	\qquad i = 0,1,\ldots,n.
\end{eqnarray}

In the same way, for the singular part of integral $I_{n+1}$ one obtains
\begin{equation}
\label{In+1sing}\fl
	I^{\textrm{sing}}_{n+1}
	= a\cdot F^{\textrm{sing}}(t_{n+1}\mid \Delta-\sum_{j=0}^{n}t_{j})
	F(t_{n}\mid \Delta-\sum_{j=0}^{n-1}t_{j})\ldots
	F(t_{1}\mid \Delta-t_{0}) F(t_{0}\mid \Delta),
\end{equation}
where $a$ is the $\delta$-function's mass in $f(s)$, see (\ref{fsing}).

Taking into account (\ref{P(t|s)sing}), (\ref{Iising}) and (\ref{In+1sing}), for the singular part of the probability density $P(t_{n+1},\ldots,t_{0})$ one obtains
\begin{eqnarray}
\fl
	P^{\textrm{sing}}(t_{n+1},t_{n},\ldots,t_{0}) &=
	\sum_{i=0}^{n+1} I_{i}^{\textrm{sing}} \nonumber
\\\fl\nonumber
	&=\sum_{i=0}^{n}A_{i}\cdot\delta\Big(\sum_{j=i+1}^{n+1}t_{j}-\Delta\Big)
	+A_{n+1}\cdot \delta(t_{n+1}+&\ldots+t_{0}-\Delta),
\\\fl
\label{Psing}
	&\qquad &\sum_{i=0}^{n}t_{i}<\Delta,
\end{eqnarray}
where $A_{i}$ and $A_{n+1}$ denote regular factors, defined by the following expressions: 
\begin{eqnarray}
\fl\nonumber
	 A_{i} &= 
	\lambda t_{n+1}\ \rme^{-\lambda t_{n+1}}\cdot
	F(t_{n}\mid \Delta - \sum_{j=i+1}^{n-1}t_{j})\ldots 
	F(t_{i+2}\mid \Delta-t_{i+1})  F(t_{i+1}&\mid \Delta)
\\\nonumber\fl
	&\times\int_{\sum_{j=0}^{i-1}t_{j}}^{\sum_{j=0}^{i}t_{j}} 
	F(t_{i}\mid s_{0}-\sum_{j=0}^{i-1}t_{j})
	F(t_{i-1}\mid s_{0}-\sum_{j=0}^{i-2}t_{j})\ldots
	F(t_{1}\mid s_{0}-t_{0}) &F(t_{0}\mid s_{0}) f(s_{0}) \rmd s_{0},
\\\fl\label{Ai}
	&\qquad &i = 0,1,\ldots,n,
\end{eqnarray}
\begin{equation}
\label{An+1}
\fl
	A_{n+1} = a\cdot\lambda t_{n+1}\ \rme^{-\lambda t_{n+1}}\cdot
	F(t_{n}\mid \Delta-\sum_{j=0}^{n-1}t_{j})\ldots
	F(t_{1}\mid \Delta-t_{0}) F(t_{0}\mid \Delta).
\end{equation}
Obviously, each factor $A_{i}$, $i=0,\ldots,n$, gives the probability to obtain $(n+1-i)$ successive output ISIs of overall duration exactly equal $\Delta$. 
And $A_{n+1}$
 gives the probability to obtain $(n+2)$ successive output ISIs of overall duration exactly equal~$\Delta$.

The presence of $\delta$-functions in joint probability density $P(t_{n+1},\ldots,t_{0})$ can be additionally
explained as follows. If at the beginning of $(i+1)$-th ISI, the impulse enters the line, then output interval $t_{n+1}$ will start with that same impulse in the feedback line with time to live equal $s_{n+1} = \Delta-\sum_{j=i+1}^{n}t_{j}$. To trigger BN after time exactly equal $s_{n+1}$ after that, it is nesessary to obtain one impulse from the Poisson stream during time interval $s_{n+1}$. This event has non-zero probability, therefore we have the non-zero probability of an output ISI exactly equal to $s_{n+1}$: $t_{n+1}=\Delta-\sum_{j=i+1}^{n}t_{j}$. This gives the corresponding $\delta$-functions in ISI probability density. The term with $\delta(t_{n+1}+\ldots+t_{0}-\Delta)$ corresponds to the case, when the impulse enters the line at the beginning of $t_{0}$.

From (\ref{defcond}) 
and (\ref{Psing}) one can easily derive the following expression for the conditional probability density:
\begin{eqnarray}
\fl
	P^{\textrm{sing}}(t_{n+1}\mid t_{n},\ldots,t_{0})
	&= \frac{1}{P(t_{n},\ldots,t_{0})}
	\sum_{i=0}^{n}A_{i}\cdot\delta\Big(\sum_{j=i+1}^{n+1}t_{j}-\Delta\Big)\nonumber
\\
\label{Pcondsing}
	&+ \frac{A_{n+1}}{P(t_{n},\ldots,t_{0})}\cdot \delta(t_{n+1}+\ldots+t_{0}-\Delta),
	\qquad \sum_{i=0}^{n}t_{i}<\Delta,
\end{eqnarray}
where $A_{i}$ and $A_{n+1}$ are defined in (\ref{Ai}) and (\ref{An+1}). It should be outlined, that joint probability density $P(t_{n},\ldots,t_{0})$ has no singularities at the domain $t_{n} < \Delta - \sum_{i=0}^{n-1} t_{i}$, see (\ref{Psing}) with $(n-1)$
 substituted instead of $n$. 

As one can see, function $P(t_{n+1}\mid t_{n},\ldots,t_{0})$ contains singularty at $t_{n+1} = \Delta - t_{n}-t_{n-1}-\ldots-t_{0}$. The dependence of the singular part of function $P(t_{n+1}\mid t_{n},\ldots,t_{0})$ on $t_{0}$ cannot be compensated by any regular summands, therefore, the whole conditional probability density $P(t_{n+1}\mid t_{n},\ldots,t_{0})$ depends on $t_{0}$. It means, 
that the condition (\ref{def}) does not hold for any $n$ for the output stream of BN with delayed feedback. The Theorem 1 
is proven.


\section{Particular cases}
\label{sec:cases}
In previous section, we have prooven
the impossibility to represent the stream of output ISI durations for BN with delayed feedback as a 
Markov chain of any finite order. In particular, output ISI stream is neither a sequence of independent random variables, and therefore is non-renewal, nor 
it is the first-order Markovian process.

In the course of proving Theorem 1 
(see Sections~\ref{sec:outline} and \ref{sec:main}), 
we have obtained the expression for $P(t_{n+1}\mid t_{n},\ldots,t_{0})$ at the domain $\sum_{i=0}^{n}t_{i}<\Delta$ in general case of an arbitrary $n$,
see \eref{Pcond}.

In this section, we consider the two particular cases of $P(t_{n+1}\mid t_{n},\ldots,t_{0})$ when $n=0$ and $n=1$, namely, the single-moment conditional probability density $P(t_{1}\mid t_{0})$ and the double-moment conditional probability density $P(t_{2}\mid t_{1},t_{0})$ and obtain the expressions for $P(t_{1}\mid t_{0})$ and $P(t_{2}\mid t_{1},t_{0})$ for domain (\ref{domain}), as well 
as for all other possible domains, which were omitted in general consideration.

\subsection{Conditional probability density $P(t_{1}\mid t_{0})$}
\label{sec:corr}

In order to derive the exact expression for conditional probability density $P(t_{1}\mid t_{0})$ 
for neighbouring ISI durations, we take Steps 1--3, outlined in Section~\ref{sec:outline}, for $n=0$. In the case of $P(t_{1}\mid t_{0})$, there are only three domains, on which the expressions should be obtained separately, namely cases $t_{0}<\Delta$, $t_{0}>\Delta$ and $t_0=\Delta$. 
Performing intergation in (\ref{joint}), one obtains the following expressions for $P(t_{1},t_{0})$ at these domains:
\begin{eqnarray}
	P(t_{1},t_{0}) &= F(t_{1}\mid \Delta)P(t_{0}),
	\qquad& t_{0}\ge\Delta,
\\
\nonumber
	&= F(t_{1}\mid \Delta) \int_{0}^{t_{0}}F(t_{0}\mid s_{0})f(s_{0})\rmd s_{0}
\\
	&+ \int_{t_{0}}^{\Delta}F(t_{1}\mid s_{0}-t_{0}) F(t_{0}\mid s_{0})f(s_{0})\rmd s_{0},
	\qquad	& t_{0}<\Delta.
\end{eqnarray}

Then, by definition of conditional probability densities, one obtains:
\begin{eqnarray}
\label{Ptt1}
	P(t_{1}\mid t_{0}) &= F(t_{1}\mid \Delta),
	\qquad& t_{0}>\Delta,
\\
\nonumber
	&= 
	\frac{1}{P(t_{0})}\ \Big(
	F(t_{1}\mid \Delta) \int_{0}^{t_{0}}F(t_{0}\mid s_{0})f(s_{0})\rmd s_{0}
\\
\label{Ptt}
	&+ \int_{t_{0}}^{\Delta}F(t_{1}\mid s_{0}-t_{0}) F(t_{0}\mid s_{0})f(s_{0})\rmd s_{0}\Big),
	\qquad  &t_{0}<\Delta.
\end{eqnarray}
It should be outlined, that the output ISI probability density $P(t_{0})$ has no singularities at the domain $t_{0}<\Delta$. 
Indeed, due to (\ref{P(t|s)sing})--(\ref{P(t)}), the only $\delta$-function contained in $P(t_0)$ is placed at $t_0=\Delta$, see Figure~{\ref{fig:Ptfs}}~(a).

In vicinity of the point $t_0=\Delta$, the single-moment conditional probability density can be derived as
\begin{equation}
\label{Pttd}
	P(t_1\mid t_0=\Delta) =
	\lim_{\epsilon \to 0} 
	\frac{\int\limits_{\Delta-\epsilon}^{\Delta+\epsilon}\rmd t_0 P(t_1,t_0)}
	{\int\limits_{\Delta-\epsilon}^{\Delta+\epsilon}\rmd t_0 P(t_0)},
\end{equation}
which just gives $\delta$-functions' masses both in numerator and denominator, and delivers
\begin{equation}
\label{Ptt2}
	P(t_{1}\mid t_{0}) = F(t_{1}\mid \Delta),
	\qquad t_{0}=\Delta.
\end{equation}

Expressions (\ref{Ptt1}), \eref{Ptt} and (\ref{Ptt2}) can be understood as follows. Since $t_{0}\ge\Delta$, one can be sure that the line has time to get free from impulse during $t_{0}$, therefore at the moment of next firing (at the beginning of $t_{1}$) the impulse enters the line and has time to live equal $\Delta$. In the case of $t_{0}<\Delta$, see (\ref{Ptt}), two possibilities arise. The first term corresponds to the scenario, when the feedback line discharges conveyed impulse within time interval $t_{0}$, and the second one represents the case when at the beginning of $t_{1}$ the line still keeps the same impulse as at the beginning of $t_{0}$. 

It can be shown, that the following normalization conditions take place:\\
$\int\limits_{0}^{\infty}\rmd t_1 P(t_1\mid t_0)=1$, and
$\int\limits_{0}^{\infty}\rmd t_0 P(t_1, t_0)=P(t_1)$.

The singular part of $P(t_{1}\mid t_{0})$ can be easily extracted:
\begin{eqnarray}
\label{Pttsing1}
	P^{\textrm{sing}}(t_{1}\mid t_{0}) &= 
	\rme^{-\ld}\ld\cdot\delta(t_{1}-\Delta),
	\qquad &t_{0}\ge\Delta,
\\\nonumber
	&=\frac{\lambda t_{1}\ \rme^{-\lambda t_{1}}}{P(t_{0})}\ \Big(
	\int_{0}^{t_{0}}F(t_{0}\mid s_{0})f(s_{0})\rmd s_{0} \cdot\delta(t_{1}-\Delta&)+
\\
\label{Pttsing2}
	&+ a\ F(t_{0}\mid \Delta)\cdot\delta(t_{0}+t_{1}-\Delta) \Big),
	\qquad  &t_{0}<\Delta.
\end{eqnarray}
Obviously, expression \eref{Pttsing2} could be obtained directly from \eref{Ai}--\eref{Pcondsing} by substituting $n=0$.

As it can be seen from (\ref{Pttsing1}) and (\ref{Pttsing2}), the number 
of $\delta$-functions in $P(t_{1}\mid t_{0})$ and their positions  depend on $t_{0}$, therefore the conditional probability density $P(t_{1}\mid t_{0})$ cannot be reduced to output ISI probability density $P(t_{1})$. Therefore, the neihgbouring output ISIs of BN with delayed feedback are correlated, as expected.

\begin{figure}
	\includegraphics[width=0.5\textwidth,angle=0]{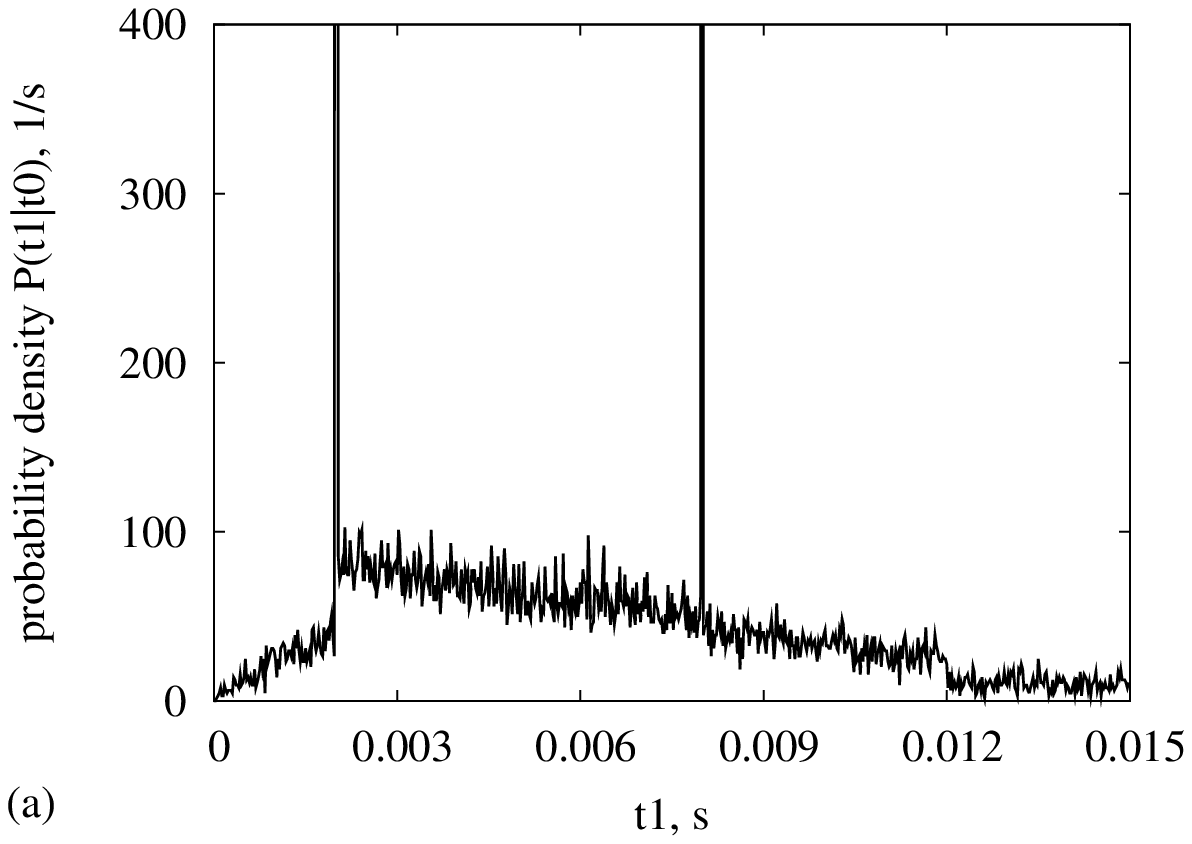}
	\includegraphics[width=0.5\textwidth,angle=0]{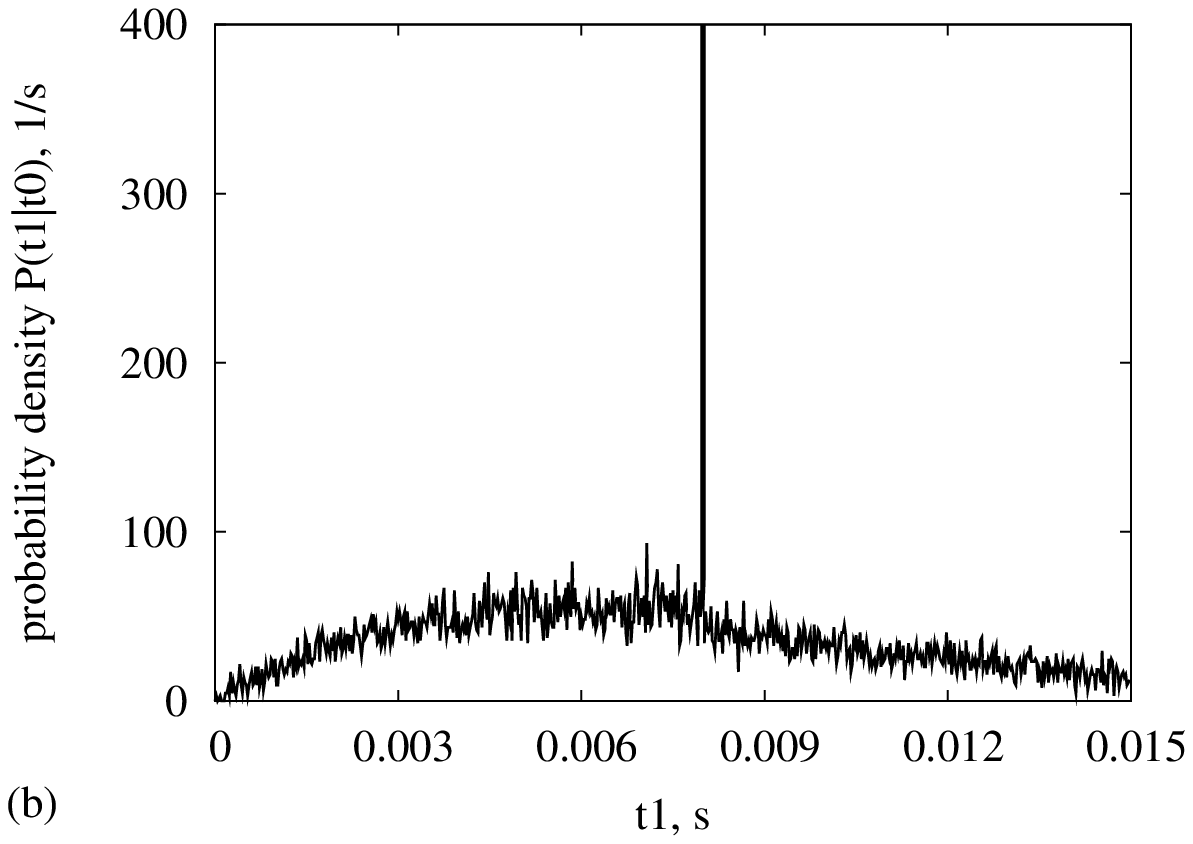}
\caption{Conditional probability density $P(t_{1}\mid t_{0})$ for $\tau$ = 10 ms, $\Delta$ = 8 ms, $\lambda$ = 150 s$^{-1}$, $N_0=2$, $t_{0}$=6 ms (a) and $t_{0}$= 11 ms (b), found numerically by means of Monte-Carlo method (the number of firings accounted $N=30\,000$).}
\label{fig:Ptt}       
\end{figure}

Examples of $P(t_{1}\mid t_{0})$, found for two domains numerically, by
means of Monte-Carlo method (see Section~\ref{sec:num} for details),
 are placed at Figure~\ref{fig:Ptt}.

\subsection{Conditional probability density $P(t_{2}\mid t_{1},t_{0})$}
\label{sec:mark}
In order to derive the exact expression for conditional probability density $P(t_{2}\mid t_{1},t_{0})$ for the succecive ISI durations, we take Steps 1--3, outlined in 
Section~\ref{sec:outline}, for $n=1$. In the case of $P(t_{2},t_{1},t_{0})$, there are 
six domains, on which the expressions should be obtained separately, namely,
the domain
\begin{equation*}
	D_1=\{t_1,t_0\mid t_1+t_0<\Delta\},
\end{equation*}
which was already utilized in Section~\ref{sec:main}, and five remaining: 
\begin{eqnarray}
\nonumber
	D_2&=\{t_1,t_0\mid\qquad t_0\ge\Delta \qquad\textrm{and}\qquad t_1\ge\Delta\},
\\\nonumber
	D_3&=\{t_1,t_0\mid\qquad t_0<\Delta \qquad\textrm{and}\qquad t_1\ge\Delta\},
\\\nonumber
	D_4&=\{t_1,t_0\mid\qquad t_0\ge\Delta \qquad\textrm{and}\qquad t_1<\Delta\},
\\\nonumber
	D_5&=\{t_1,t_0\mid\qquad t_0<\Delta \qquad\textrm{and}\qquad \Delta-t_0 < t_1 < \Delta\},
\\\nonumber
	d&=\{t_1,t_0\mid\qquad t_{0}+t_{1}=\Delta\}.
\end{eqnarray}
In the case, when the exact equality $t_{0}+t_{1}=\Delta$ holds, namely, if $(t_1,t_0)\in d$, the product $P(t_2\mid t_1,t_0)\rmd t_2$ gives the probability to obtain an output ISI of duration within interval $[t_2;t_2+\rmd t_2[$, provided the overall duration of two previous ISIs accurately equals $\Delta$.

Expressions for $P(t_2\mid t_1,t_0)$ can be found exactly on each domain:
\begin{eqnarray}
\fl
	P(t_{2}\mid t_{1},t_{0}) &= F(t_{2}\mid \Delta),
	\qquad	&(t_0, t_1)\in D_2,
\\
\fl
	&= F(t_{2}\mid \Delta),
	\qquad	&(t_{0}, t_1) \in D_3,
\\
\label{Ptttd}
\fl
	&= F(t_{2}\mid \Delta),
	\qquad	&(t_{0}, t_1) \in d,
\\
\fl
	&= F(t_{2}\mid \Delta-t_{1}),
	\qquad	&(t_{0}, t_1) \in D_4,
\\
\label{PtttD5}
\fl\nonumber
	&= \frac{1}{P(t_{1},t_{0})}\ \Big(
	F(t_{2}\mid \Delta-t_{1}) F(t_{1}\mid \Delta) 
	\int_{0}^{t_{0}}F(t_{0}\mid s_{0})f(s_{0})&\rmd s_{0}
\\
\fl
	&+ F(t_{2}| \Delta)
	\int_{t_{0}}^{\Delta}F(t_{1}| s_{0}-t_{0})F(t_{0}| s_{0})f(s_{0})\rmd s_{0}\Big),
	\qquad	
	&(t_{0}, t_1) \in D_5,
\\
\fl\nonumber
	&= \frac{1}{P(t_{1},t_{0})}\ \Big(
	F(t_{2}\mid \Delta-t_{1}) F(t_{1}\mid \Delta) 
	\int_{0}^{t_{0}}F(t_{0}\mid s_{0})f(s_{0})&\rmd s_{0}
\\\fl\nonumber
	&+ F(t_{2}\mid \Delta)\int_{t_{0}}^{t_{0}+t_{1}}
	F(t_{1}\mid s_{0}-t_{0})F(t_{0}\mid s_{0})f(s_{0})\rmd s_{0}
\\\fl\nonumber
	&+\int_{t_{0}+t_{1}}^{\Delta} F(t_{2}| s_{0}-t_{0}-t_{1})
	F(t_{1}| s_{0}-t_{0})F(t_{0}| s_{0})f(s_{0})\rmd s_{0}\Big)&,
\\\fl
\label{PtttD1}
	&\qquad	&(t_{0}, t_{1}) \in D_1.
\end{eqnarray}
where $P(t_{1},t_{0})=F(t_{1}\mid \Delta) \int_{0}^{t_{0}}F(t_{0}\mid s_{0})f(s_{0})\rmd s_{0} + \int_{t_{0}}^{\Delta}F(t_{1}\mid s_{0}-t_{0})F(t_{0}\mid s_{0})f(s_{0})\rmd s_{0}$, according to (\ref{Ptt}). 

The probability density $P(t_{1},t_{0})$ contains $\delta$--function at the domain $d$, see \eref{Pttsing2}. In \eref{Ptttd}, the two-time conditional probability density was derived as
\begin{equation*}
	P(t_2\mid t_1,t_0) =
	\lim_{\epsilon \to 0} 
	\frac{\int\limits_{\Delta-t_0-\epsilon}^{\Delta-t_0+\epsilon}\rmd t_1 P(t_2,t_1,t_0)}
	{\int\limits_{\Delta-t_0-\epsilon}^{\Delta-t_0+\epsilon}\rmd t_1 P(t_1,t_0)},
	\qquad (t_0,t_1)\in d,
\end{equation*}
compare with \eref{Pttd}.

It is worth to notice, that $P(t_{1},t_{0})$ is regular function on both $D_{1}$ and $D_{5}$, see (\ref{PtttD5}) and (\ref{PtttD1}).
Indeed, from (\ref{Pttsing1}) and (\ref{Pttsing2}) one can see, that $P(t_{1},t_{0})$ may include singularities only at the points $t_{1}=\Delta$ and $t_{1}=\Delta-t_{0}$. None of these points belongs to $D_{1}$, or $D_{5}$.

It can be shown, that the following normalization conditions take place:\\
$\int\limits_{0}^{\infty}\rmd t_2 P(t_2\mid t_1,t_0)=1$, and
$\int\limits_{0}^{\infty}\rmd t_0 P(t_2, t_1, t_0)=P(t_2,t_1)$.

\begin{figure}
	\includegraphics[width=0.5\textwidth,angle=0]{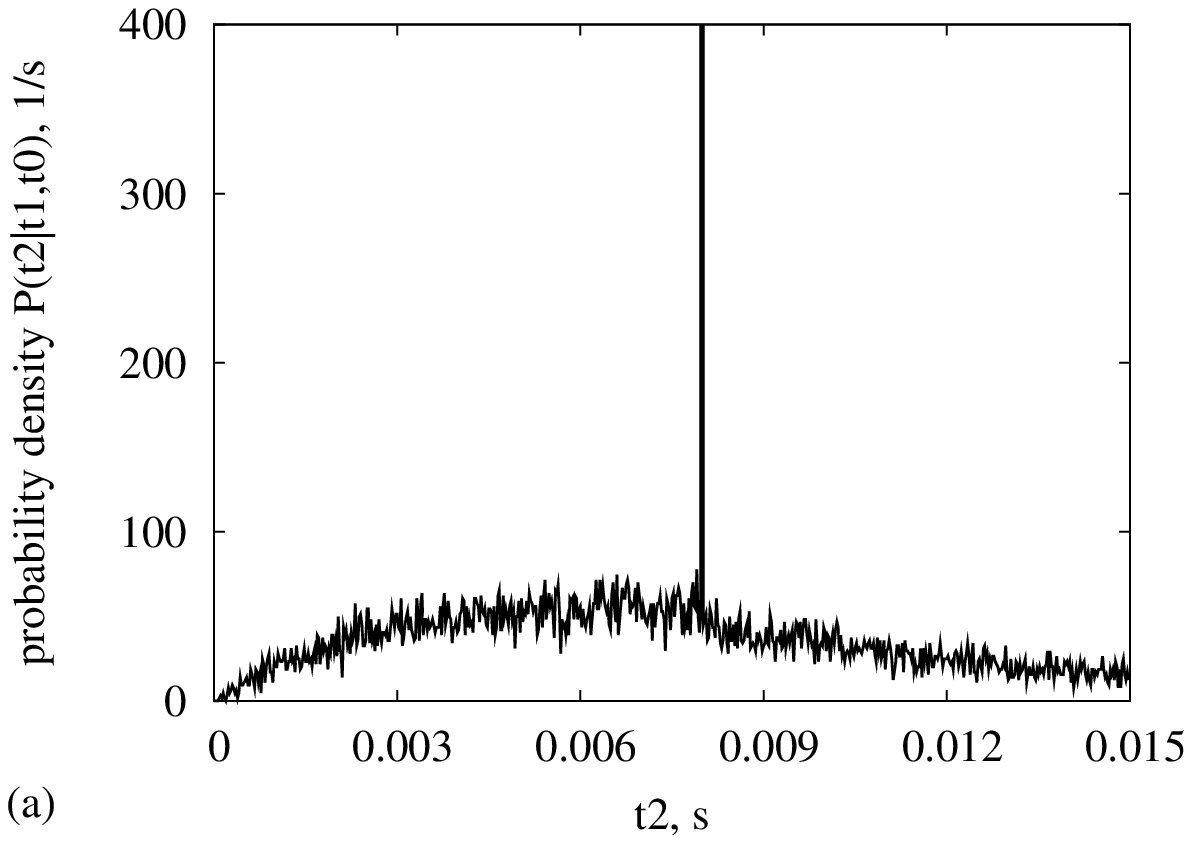}
	\includegraphics[width=0.5\textwidth,angle=0]{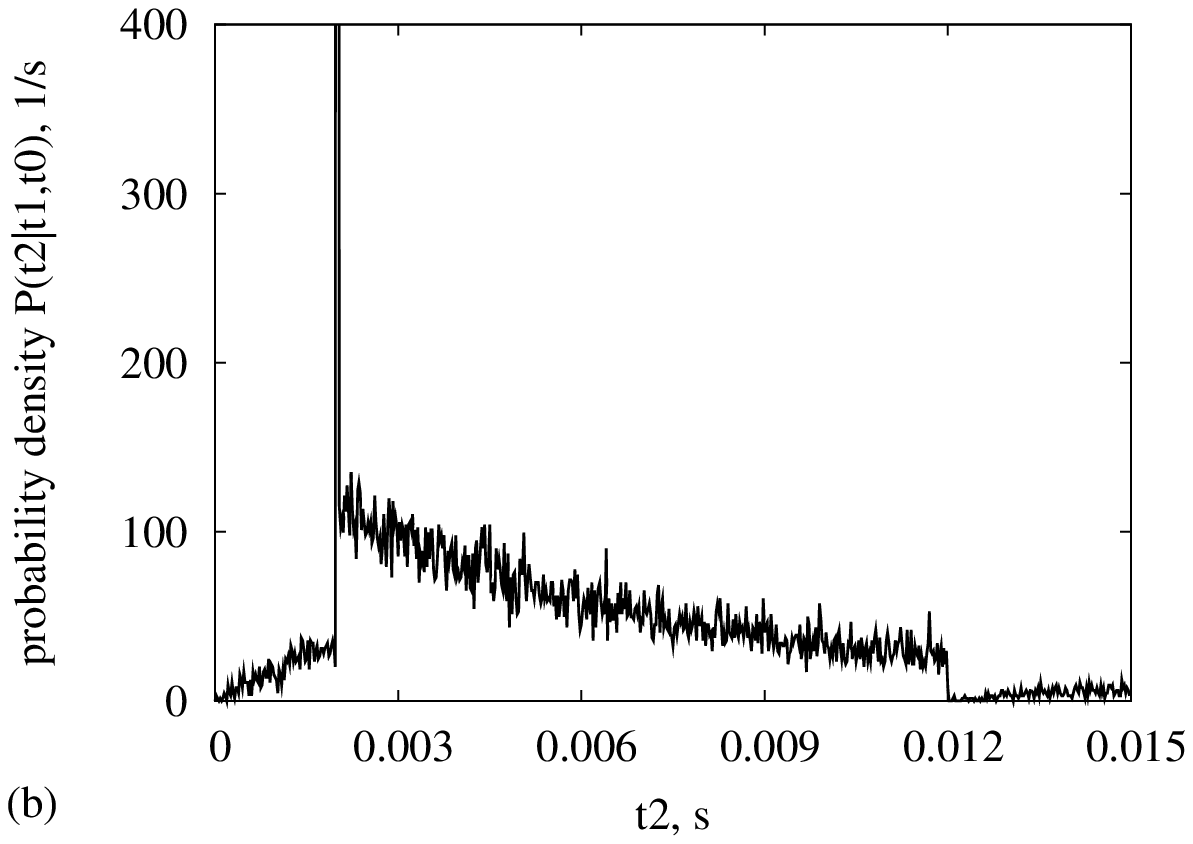}
\caption{Conditional probability density $P(t_{2}\mid t_{1}, t_0)$ for $\tau$ = 10 ms, $\Delta$ = 8 ms, $\lambda$ = 150 s$^{-1}$, $N_0=2$, $t_{1}$=13 ms, $t_{0}$=13 ms (a) and $t_{1}$ = 6 ms, $t_{0}$ = 13 ms (b), found numerically by means of Monte-Carlo method ($N=30\,000$).}
\label{fig:Pttt1}       
\end{figure}

The singular part of the conditional probability density $P(t_{2}\mid t_{1}, t_0)$ can be derived as follows:
\begin{eqnarray}
\label{Ptttsing1}
\fl
	P^{\textrm{sing}}(t_{2}\mid t_{1},t_{0}) 
	&= \rme^{-\lambda t_2}\lambda t_2\cdot\delta(t_{2}-\Delta),
	\qquad	&(t_{0}, t_{1}) \in D_2\cup D_3\cup d,
\\\fl
	&= \rme^{-\lambda t_2}\lambda t_2\cdot\delta(t_{1}+t_{2}-\Delta),
	\qquad	&(t_{0}, t_{1}) \in D_4.
\\
\fl\nonumber
	&=\frac{\rme^{-\lambda t_2}\lambda t_2}{P(t_{1},t_{0})}
	\cdot \Big(F(t_{1}\mid \Delta) 
	\int_{0}^{t_{0}}F(t_{0}\mid s_{0})f(s_{0})&\rmd s_{0}
	\cdot\delta(t_{1}+t_{2}-\Delta)
\\
\fl\nonumber
	&+ \int_{t_{0}}^{\Delta}F(t_{1}\mid s_{0}-t_{0})F(t_{0}\mid s_{0})f(s_{0})\rmd s_{0}
	\cdot\delta(t&_{2}-\Delta)\Big)
\\
\fl
	& \qquad&(t_{0}, t_{1}) \in D_5,
\\
\fl\nonumber
	&=\frac{\rme^{-\lambda t_{2}}\lambda t_{2}}{P(t_{1},t_{0})}
	\Big(\int_{t_{0}}^{t_{0}+t_{1}}
	F(t_{1}| s_{0}-t_{0})F(t_{0}| s_{0})f(&s_{0})\rmd s_{0} 
	\cdot\delta(t_{2}-\Delta)
\\
\fl\nonumber
	&+ F(t_{1}\mid \Delta) 
	\int_{0}^{t_{0}}F(t_{0}\mid s_{0})f(s_{0})\rmd s_{0}
	\cdot\delta(t_{1}+t&_{2}-\Delta)
\\
\fl\nonumber
	&+a\cdot F(t_{1}\mid \Delta-t_{0})F(t_{0}\mid \Delta)
	\cdot\delta(t_{0}+t_{1}+t_{2}&-\Delta)
	\Big),
\\\fl
\label{Ptttsing5}
	&\qquad	&(t_{0}, t_{1}) \in D_1.
\end{eqnarray}
Obviously, expression \eref{Ptttsing5} could be obtained directly from \eref{Ai}--\eref{Pcondsing} by substituting $n=1$.

As one can see, the singular part of $P(t_{2}\mid t_{1},t_{0})$ depends on $t_{0}$, therefore $P(t_{2}\mid t_{1},t_{0})$ cannot be reduced to $P(t_{2}\mid t_{1})$, which means that the output stream is not first-order Markovian.

\begin{figure}
	\includegraphics[width=0.5\textwidth,angle=0]{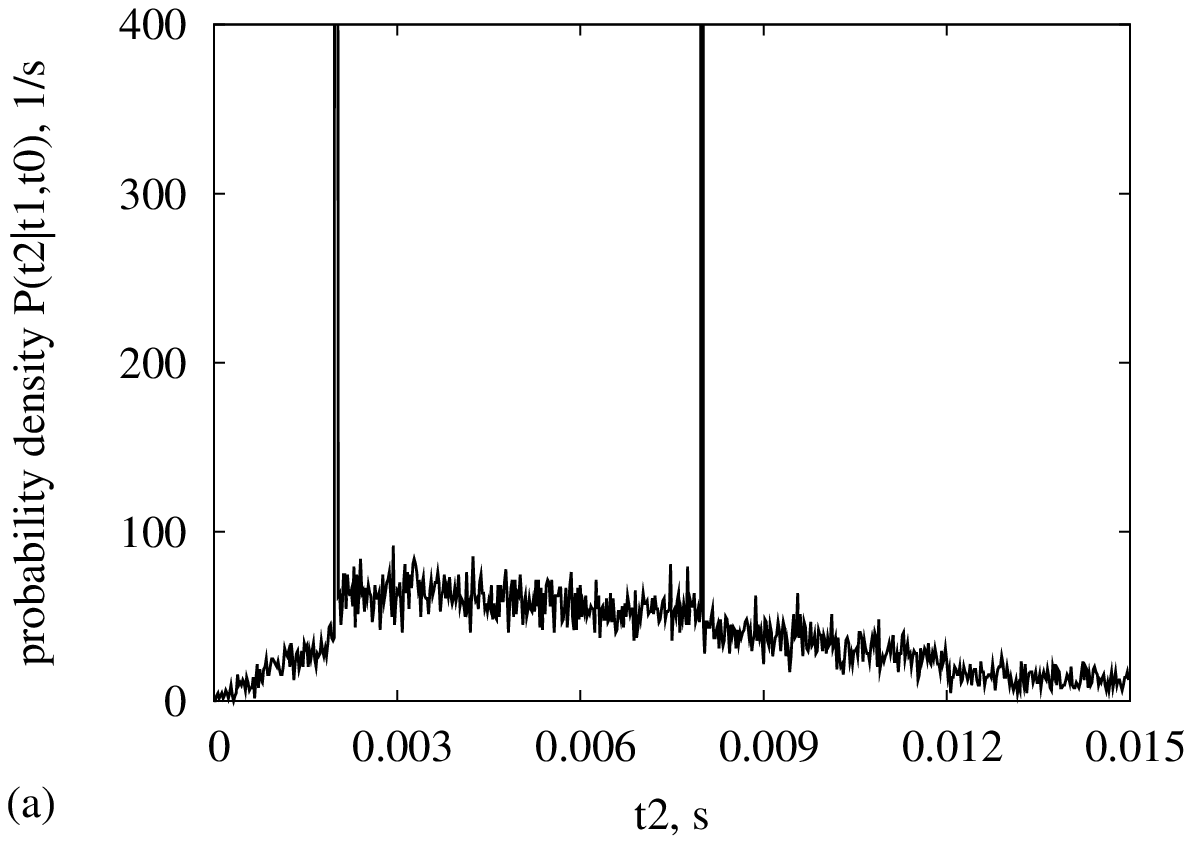}
	\includegraphics[width=0.5\textwidth,angle=0]{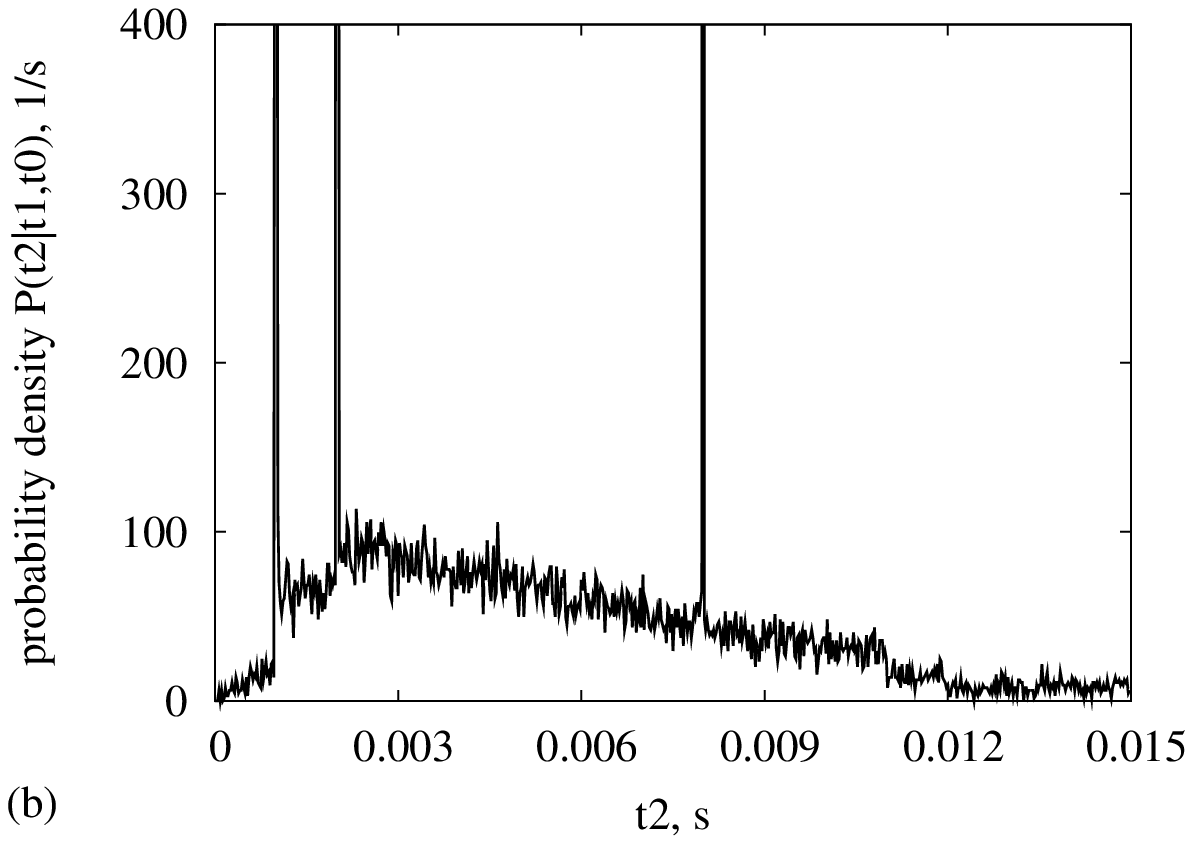}
\caption{Conditional probability density $P(t_{2}\mid t_{1}, t_0)$ for $\tau$ = 10 ms, $\Delta$ = 8 ms, $\lambda$ = 150 s$^{-1}$, $N_0=2$, $t_{1}$=6 ms, $t_{0}$=3 ms (a) and $t_{1}$ = 6 ms, $t_{0}$ = 1 ms (b), found numerically by means of Monte-Carlo method ($N=30\,000$).}
\label{fig:Pttt2}       
\end{figure}

Examples of $P(t_{2}\mid t_{1},t_0)$, found numerically for different domains, are placed at Figures~\ref{fig:Pttt1} and \ref{fig:Pttt2}.

\section{Numerical simulation}
\label{sec:num}

In order to check the correctness of obtained analytical expressions,
and also
to investigate wheather the output ISIs stream is non-Markovian
for BN with higher thresholds as well as for $N_0=2$,
numerical simulations were performed. A C++ program, containing class, which models the operation manner of BN with delayed feedback, was developed. Object of this class receives the sequence of pseudorandom numbers with Poisson probability density to its
input. The required sequences were generated by means of utilities from the GNU 
Scientific Library\footnote{http://www.gnu.org/software/gsl/}
with the Mersenne Twister generator as source of pseudorandom numbers.

Program contains function, the time engine, which brings system to the moment just before the next input signal, bypassing moments, when neither external Poisson impulse, nor impulse from the feedback line comes. So, only the essential discret events are accounted. It allows one to make exact calculations faster as compared to
 the algorithm where time advances gradually by adding small timesteps.

\begin{figure}
	\includegraphics[width=0.5\textwidth,angle=0]{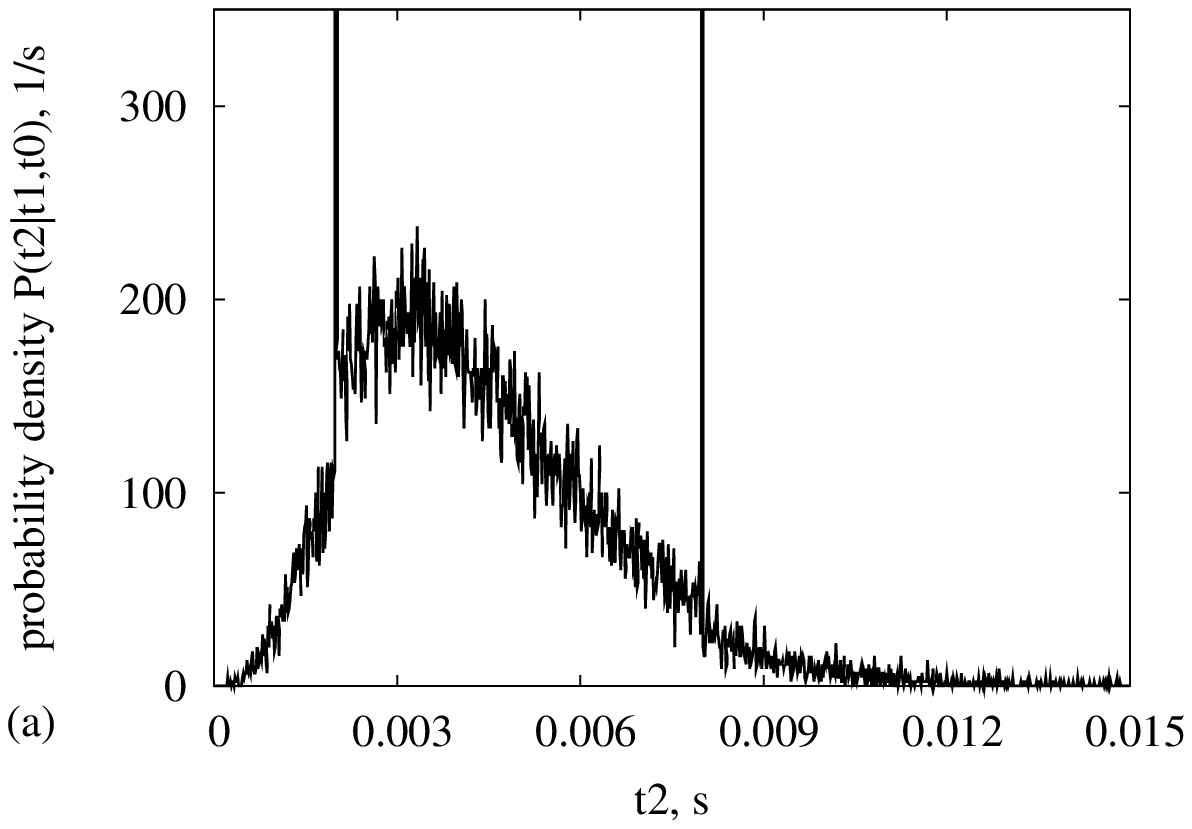}
	\includegraphics[width=0.5\textwidth,angle=0]{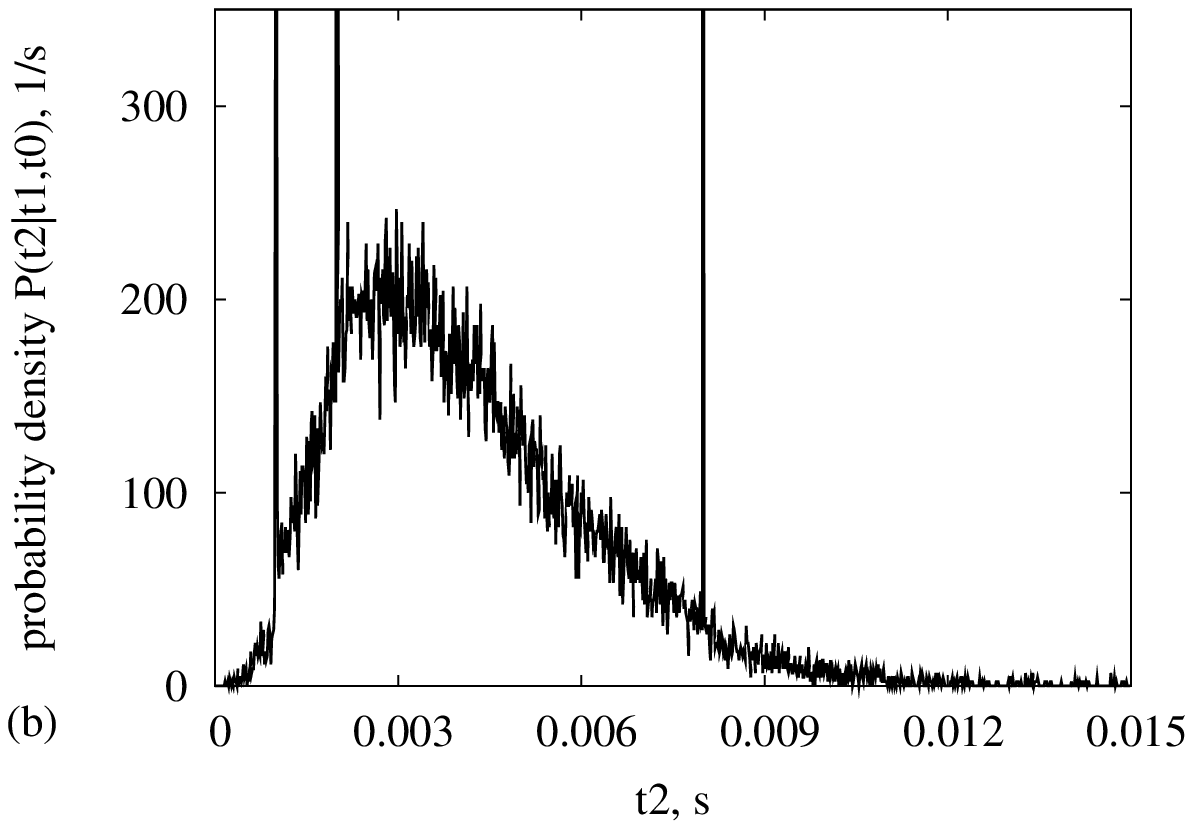}
\caption{Conditional probability density $P(t_{2}\mid t_{1}, t_0)$ for $\tau$ = 10 ms, $\Delta$ = 8 ms, $\lambda$ = 800 s$^{-1}$, $N_0=4$, $t_{1}$=6 ms, $t_{0}$=3 ms (a) and $t_{1}$ = 6 ms, $t_{0}$ = 1 ms (b), found numerically by means of Monte-Carlo method ($N=30\,000$).}
\label{fig:PtttN04}       
\end{figure}

The conditional probability densities, $P(t_{1}\mid t_{0})$
and $P(t_{2}\mid t_{1},t_{0})$, are found by counting the number of output ISI of different durations and normalization (see Figures~\ref{fig:Ptt} -- \ref{fig:PtttN04}). 
Obviously, for calculation of conditional distiributions only those ISIs are selected, which follow one or two ISIs of fixed duration,  $t_{0}$ for $P(t_{1}\mid t_{0})$ and $\{t_{1}, t_{0}\}$ for $P(t_{2}\mid t_{1},t_{0})$. The quantity, the position and the mass
of delta-functions, obtained in numerical experiments for BN with threshold 2,
coincide with those predicted analitycally in (\ref{Pttsing1}), (\ref{Pttsing2}) and (\ref{Ptttsing1}) -- (\ref{Ptttsing5}).

For $N_0>2$, conditional probability densities $P(t_{1}\mid t_{0})$
and $P(t_{2}\mid t_{1},t_{0})$ are similar to those, found for $N_0$=2.
In particular, both the quantity and position of delta-functions
coincide with those obtained for BN with threshold 2,
as expected, compare Figures~\ref{fig:PtttN04} and \ref{fig:Pttt2}.


\section{Conclusions and discussion}
\label{sec:disc}

Our results reveal the influence of the delayed feedback presence on the neuronal firing statistics.
In contrast to the cases of BN without feedback \cite{VidBiosys} and BN with instantaneous feedback \cite{BNF}, the nighbouring output ISIs of BN with delayed feedback are mutually correlated. It means that even in the simplest possible recurrent network the ISI stream cannot be treated as the renewal one. The presence of nearest ISIs correlation was reported for spike trains of a neurons in different CNS and peripheral NS structures \cite{Farkhooi,Nawrot07}. 

Moreover, we prove, that the output ISI stream of BN with delayed feedback cannot be represented as the Markov chain of any finite order. This is in accordance with rare attempts of experimental estimation of the Markov 
order of neuronal spike trains (see, e.g. \cite{RatnamNelson}, where it is established that the order, if any, must be greater than 3).

 We expect the same non-markovian property for firing statistics of any single
 neuron with delayed feedback, whatever neuronal model is used, 
 and conclude that it is namely the delayed feedback presence results in non-markovian statistics found.
One should take this fact into account during analysis of neuronal spike trains obtained from any recurent network.

\end{document}